\documentclass[12pt]{article}

\usepackage[dvips]{graphicx}
\usepackage {amsfonts}
\usepackage {amssymb}
\usepackage { pgfpages }

\begin{document}

\begin{center}
{\bfseries  DEUTERON-PROTON BACKWARD ELASTIC SCATTERING AT
GeV ENERGIES}

\vskip 5mm

Nadezhda Ladygina$^\dag$

\vskip 5mm

{\small
 {\it
Laboratory of High Energy Physics, Joint Institute for Nuclear Research, 141980 Dubna, Russia
}
\\
$\dag$ {\it
E-mail: nladygina@jinr.ru
}}
\end{center}

\vskip 5mm

\begin{center}
\begin{minipage}{150mm}
\centerline{\bf Abstract}
Deuteron-proton elastic scattering is considered at the energies from 880 MeV to 2 GeV at the scattering angles $\theta^*\ge 140^\circ$.
The multiple-scattering method is used to calculate the reaction amplitude.
Four  reaction mechanisms are taken into account: one-nucleon-exchange, single-scattering, and double-scattering
with a nucleon and a delta-isobar in the intermediate state. The method allows calculating both unpolarized differential cross section and any polarization observables. 
 All the results obtained are compared to the experimental
data.  

\end{minipage}
\end{center}

\section{Introduction}
\label{sec:intro}

It is customary to consider the deuteron-proton elastic scattering as a source of  information 
about the deuteron structure. Meanwhile, studying the reaction at intermediate energies is a good opportunity to understand how various reaction mechanisms work.
Deuteron is the simplest nucleus. Investigating the deuteron-nucleon scattering, we gain new knowledge about  both the coupled system and the nucleon-nucleon interaction.


Although  the  first  nucleon-deuteron  experiments were performed already in the 1950s 
this reaction is still the subject of investigations.
Recently new results  have
been obtained at the Nuclotron (JINR, Dubna) \cite{pl}-\cite{at}.
  It has been measured both the differential cross sections
 at the deuteron  energies
between 1000 and 1800 MeV \cite{at}
and polarization observables $A_y$, $A_{yy}$, $A_{xx}$ in a wide energy range from 400 MeV to 1800 MeV \cite{dss}.   However, despite the large amount of the data, there is no theory capable of describing them at such energies.

The results of the measurement of the tensor analyzing power $T_{20}$ and the polarization transfer from the vectorially polarized deuteron to the final proton $\varkappa_0$ at the
deuteron energies from 0.3 to 2.4 GeV at the scattering angle $\theta^*=180^\circ$
were presented in ref.\cite{punjabi}. In the case when the reaction proceeds according to the simplest one-nucleon-exchange (ONE) mechanism, the $\varkappa_0-T_{20}$ correlation forms a circle of radius $3/\sqrt{8}$. But the data do not follow the ONE circle. An attempt to explain the discrepancy between the data and theory predictions by relativistic effects was unsuccessful.
In ref.\cite{kaptari}, relativistic effects in the dp backward scattering were investigated in detail. The reaction was considered within the ONE frame. It was shown that despite the importance of  relativistic effects
 it is impossible to describe the data by  taking into account only ONE reaction mechanism.       

In previous papers we   suggested a model based on the multiple expansion of the reaction amplitude 
 in powers of the nucleon-nucleon $t$-matrix \cite{japh}-\cite{epja2009}.
We avoid the angular momentum decomposition because of the need to take into account too many partial waves to achieve convergence 
at intermediate and high energies. Instead of this, we use some parameterizations both for the nucleon-nucleon $t$-matrix and 
 for the deuteron wave function. These parameterizations were defined independently of our approach. We do not introduce
any new parameters to achieve better agreement between the calculation results and the data.
Because the energies are high enough for the manifestation of mesonic intermediate states, we include into consideration the $\Delta$-isobar term \cite{epja2016}. This modification is the more relevant because we study the dp-elastic scattering at the large scattering angles.

The basic points of the model are introduced in Sect.2. We give  definitions of the observables
of interest in Sect.3. The  results obtained are presented and discussed in Sect.4. Finally, some 
conclusions.

\section{General formalism}
\label{sec:gen}
The amplitude of the deuteron-proton elastic scattering $\cal J$
can be defined through the matrix element of the transition operator $U_{11}$:
\begin{eqnarray}
\label{U11}
U_{dp\to dp}&=&\delta(E_d+E_p-E^\prime_d-E^\prime _p) {\cal J}=
\nonumber\\
&&<1(23)|[1-P_{12}-P_{13}]U_{11}|1(23)>.
\end{eqnarray}
Here, the notation of $U_{11}$ means that the transition is between states in which  nucleon 1 is free, and nucleons 2 and 3 form the deuteron  in both the initial and final states. We denote such a state as $|1(23)>$.
Since the initial and final states are antisymmetric with respect to the permutation of two nucleons, Eq.(\ref{U11}) includes the permutation operators $P_{ij}$.

In order to calculate the three particle transition operator $U_{ij}$, Alt, Grassberger, and Sandhas \cite{AGS} suggested the following set of equations for rearrangement scattering operators:

\begin{eqnarray}
\label{AGS}
U_{11}&=&~~~~~~~~t_2g_0U_{21}+t_3g_0U_{31},
\nonumber\\
U_{21}&=&g_0^{-1}+t_1g_0U_{11}+t_3g_0U_{31},
\\
U_{31}&=&g_0^{-1}+t_1g_0U_{11}+t_2g_0U_{21},
\nonumber
\end{eqnarray}
where $t_1=t(2,3)$, etc., is the $t$-matrix of the two-nucleon interaction and
$g_0$ is the free three-particle propagator. The indices $ij$ for the
transition operators $U_{ij}$ denote free particles $i$ and $j$ in the final
 and initial states, respectively.

To define $U_{11}$ we iterate Eq.(\ref{AGS}) over the two-particle matrix $t_i$.
Restricting the sequence to
 the $t_i$-second-order terms,  we can present
the reaction amplitude as a sum of the  four contributions:
\begin{eqnarray}
\label{contrib}
{\cal J}_{dp\to dp}&=&{\cal J}_{\rm ONE}+{\cal J}_{\rm SS}+{\cal J}_{\rm DS}+{\cal J}_{\Delta},
\end{eqnarray}
This equation includes one-nucleon exchange (ONE),
single scattering (SS), double scattering (DS), and 
 rescattering with
$\Delta$ -excitation in the intermediate state.

\begin{figure}
\centering
\includegraphics[width=12cm,clip]{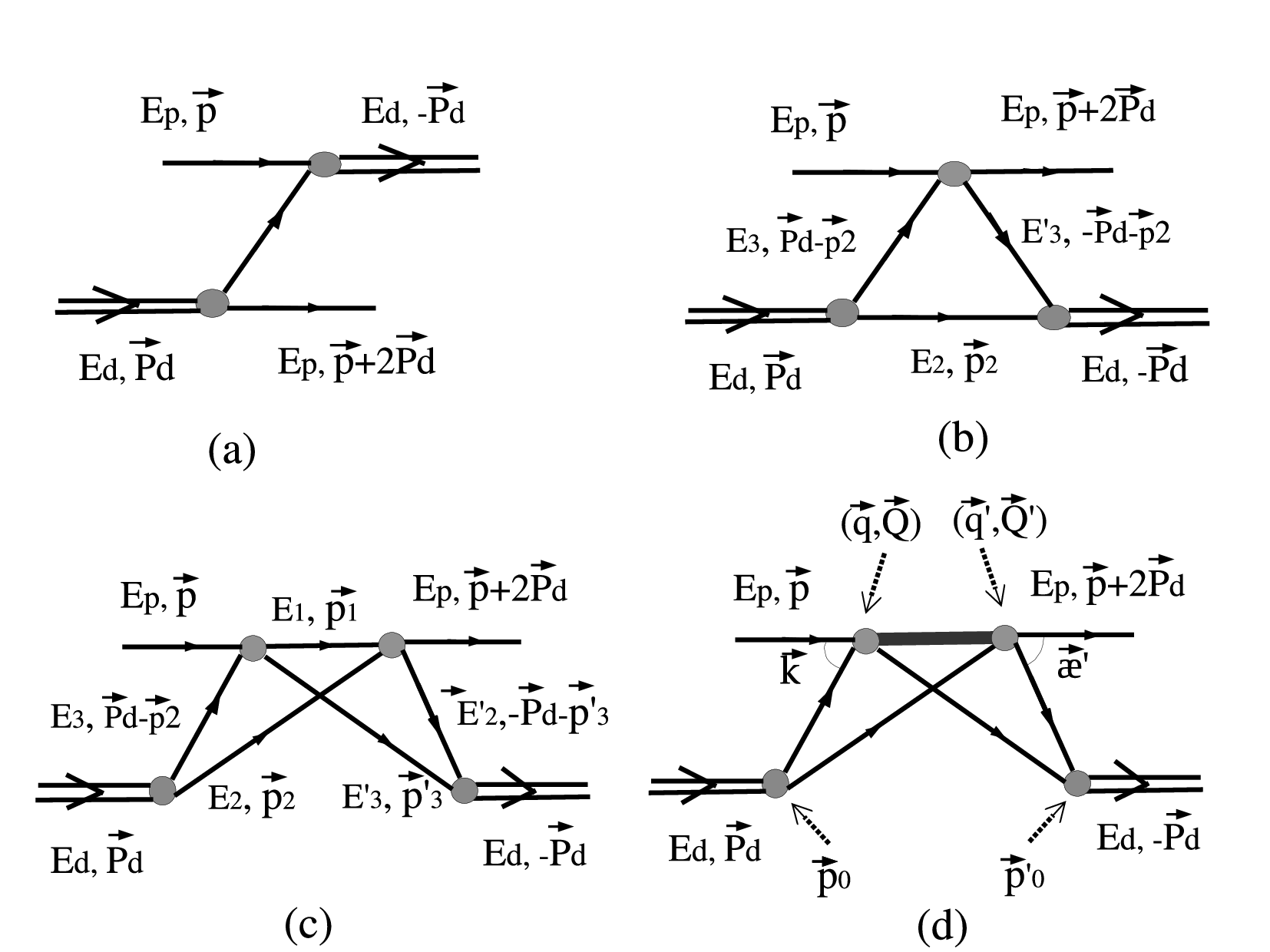}
\caption{The diagrams included into consideration:
(a) the one-nucleon exchange diagram; (b) the single
scattering diagram; (c)  the double scattering diagram
with a nucleon in the intermediate state;
(d) the double scattering diagram
with $\Delta$-isobar in the intermediate state.}
\label{diagr}       
\end{figure}

The  diagram corresponding to ONE is presented in Fig.1a.
This term does not contain the NN $t$-matrix. Here, the initial deuteron
breaks up to two nucleons, a proton and a neutron.  The neutron interacts with 
the beam proton and they form the deuteron in the final state while the proton moves as a spectator.
The contribution of ONE into the deuteron-proton reaction amplitude is described as
\begin{eqnarray}
{\cal J}_{ONE}&=&-2<1(23)|P_{12}g_0^{-1}|1(23)>.
\end{eqnarray}

Explicitly this expression can be presented in the following 
form:
\begin{eqnarray}
\label{ONE}
{\cal J}_{ONE}&=&-\frac{1}{2}(E_d-E_p-\sqrt{m_N^2+\vec p~^2-\vec P_d\-^2})
\cdot
\\
&&<\vec p^\prime m^\prime;-\vec P_d {\cal M}_d^\prime|
\Omega^\dagger_d(23)
[1+(\mbox{\boldmath$\sigma_1\sigma_2$})]
\Omega_d(23)|
\vec P_d {\cal M}_d;\vec p m >~~,
\nonumber
\end{eqnarray}
where $\Omega_d(23)$ is the deuteron wave function, the kinematic factor is from the three-
nucleon free propagator $g_0^{-1}$ and  the permutation
operator in spin space is
 $P_{12}(\sigma )=\frac{1}{2}[1+(\mbox{\boldmath$\sigma_1\sigma_2$})]$. Spin quantum numbers are denoted as $m_i$ and $ {\cal M}_d$
for nucleons and deuterons, respectively.

For our further calculations we need, first of all, an expression for the deuteron wave function.
There are well-known parameterizations of the  deuteron wave function, such as Bonn\cite{B}, CD Bonn \cite{cd}, Paris\cite{par} etc. But all these parameterizations are non-relativistic and  are functions of the relative momentum of the nucleons. Moreover, they are defined  in the deuteron rest frame. However, we have two deuterons: one in the initial and the other in the final states. In such a way, if we choose the  rest frame of the initial deuteron, the final deuteron moves with rather large momentum and the relative momentum of the nucleons is more than 1 GeV/c.
Therefore, we perform all the calculations  in the deuteron Breit frame, where
the deuterons move  in opposite directions with equal momenta (Fig.1).
\begin{eqnarray}
E_d=E^\prime_d=\sqrt{M_d^2+\vec P_d^2}, ~~~~
E_p=E^\prime_p=\sqrt{m^2+\vec p^2}, ~~~~(\vec p\vec P_d)=-\vec P_d^2.
\end{eqnarray}
It allows us to minimize  the relative momenta of the nucleons in  both
deuterons. As a consequence, the non-relativistic deuteron wave function
can be applied in the energy range under consideration.

In the rest frame the non-relativistic wave function  of the deuteron 
depends only on one variable $\vec p_0$, which is the
 relative momentum of the  outgoing proton and neutron:
\begin{eqnarray}
\label{dwf0}
&&<\mu_p \mu_n|\Omega_d|{\cal M}_d>=
\frac{1}{\sqrt{4\pi}}<\mu_p \mu_n|\{ u(p_0)+
\frac{w(p_0)}{\sqrt 8}
[3(\mbox{\boldmath$\sigma_1$} \hat p_0)(\mbox{\boldmath$\sigma_2$}\hat p_0)-(
\mbox{\boldmath $\sigma_1$ $\sigma_2$}
)]
\}|{\cal M}_d>,
\nonumber
\end{eqnarray}
where $u(p_0)$ and $w(p_0)$ describe the $S$ and $D$ components of 
the deuteron wave function   \cite{B}, \cite{cd}, \cite{par}, $\hat p_0$ is 
the unit vector in $\vec p_0$ direction.

In order to get the wave function of the moving deuteron, it is necessary to apply
the Lorenz transformations for the kinematical variables and Wigner rotations for 
the spin states. This procedure has been expounded in ref.\cite{japh}.
The proton-neutron relative momenta    for the initial
 $\vec p_0$ and final $\vec p_0^\prime $ deuterons are expressed in ONE case (Fig.1a) as: 
\begin{eqnarray}
\label{p0}
\vec p_0=\vec p +\vec P_d\left[ 1+\frac{E_n+E^*}{E_p+E_n+E^*}\right],~~~~
\vec p_0^\prime =\vec p +\vec P_d\left[ 1-\frac{E_n+E^*}{E_p+E_n+E^*}\right]~~.
\end{eqnarray}
Here $E_n=\sqrt{m_N^2+\vec p~^2-\vec P_d\-^2}$ 
 and $E^*=\sqrt{(E_p+E_n)^2-\vec P_d\-^2}/2$ 
are the struck neutron energy in the moving deuteron frame and rest deuteron
frame, respectively. Note that $|\vec p_0|=|\vec p_0^\prime |$. It can be seen from Eq.\ref{p0} that
the relative momentum of the nucleons in the moving deuteron depends on two momenta: one of the nucleon and the deuteron.
It is not surprising since a two-particle system has two variables: the relative momentum  and the total momentum 
 of the pair, or other combinations of these variables. When we are dealing with a non-relativistic wave function in the deuteron  rest frame we also have two momenta, but one of them is equal to zero.

The next term in the $dp$-elastic scattering amplitude 
Eq.(\ref {contrib}) is the single scattering one:
\begin{eqnarray}
{\cal J}_{SS}&=&2<1(23)|[1-P_{12}]t_3|1(23)>.
\end{eqnarray}

The corresponding diagram is presented in Fig.(1b). This term is the first-order term of the nucleon-nucleon $t$-matrix.
Here, the nucleon from the initial deuteron interacts with the beam proton. This interaction is described by NN $t$-matrix.
Then, one of the scattered nucleon
 combines with another  nucleon from the initial deuteron and they form the final deuteron. 
 Since the nucleons are indistinguishable 
after scattering,  a permutation operator is present in the amplitude. The operator is included in the  definition of
 the $t$-matrix during further calculation.

Following the  standard procedure we get the expression 
for the single scattering amplitude:
\begin{eqnarray}
\label{ss}
{\cal J}_{SS}&=&\int d\vec q~^\prime <-\vec P_d {\cal M}_d^\prime |
\Omega_d^\dagger|
\vec q~^\prime m^{\prime\prime}, -\vec P_d-\vec q~^\prime m_3^\prime>
\\
&&\hspace{-1cm}
<\vec p~^\prime m^\prime, -\vec P_d-\vec q~^\prime |
\frac{3}{2}t^1_{12}+\frac{1}{2}t^0_{12}|
\vec p m, \vec P_d -\vec q~^\prime m_2^\prime >
<\vec q~^\prime m^{\prime\prime},\vec P_d-\vec q~^\prime m_2^\prime|
\Omega_d|\vec P_d {\cal M}_d>~~.
\nonumber
\end{eqnarray}
The relative momenta of two nucleons for the initial and final 
deuterons for SS-case are
\begin{eqnarray}
\vec p_0=\vec q~^\prime -\vec P_d\frac{E_2+E^*}
{E_2+E_3+2E^*}~~~~
\vec p_0^\prime =\vec q~^\prime +\vec P_d
\frac{E_2+E^{\prime *}}
{E_2+E_3^\prime +2E^{\prime *}}~~,
\end{eqnarray}
where the nucleons energies $E_2$, $E_3$, $E_3^\prime $ 
in the reference frame are defined by the 
standard manner (Fig.1b)
\begin{eqnarray}
\label{e2e3}
E_2=\sqrt{m_N^2+\vec q~^{\prime 2}}~,
~~~~~~
E_3=\sqrt{m_N^2+(\vec P_d -\vec q~^\prime)^2}~,
~~~~~~
E_3^\prime =\sqrt{m_N^2+(\vec P_d +\vec q~^\prime)^2}
\end{eqnarray}
and these energies in the center-of-mass of the two nucleons
forming the initial and final deuterons
 are equal, correspondingly, to
\begin{eqnarray}
E^*=\frac{1}{2}\sqrt{(E_2+E_3)^2-\vec P_d^2}~,
~~~~~~~
E^{\prime *}=\frac{1}{2}\sqrt {(E_2+E_3^\prime )^2-\vec P_d^2}~~.
\end{eqnarray}

The nucleon-nucleon scattering is described by the antisymmetrized $t$-matrix $t_{ij}^T$. Superscript T corresponds to the
isospin of the nucleon pair.
 We use the parameterization  of this
 matrix offered by Love and Franey  \cite {LF}. This is the on-shell NN $t$-matrix
  defined in the center-of-mass:
\begin{eqnarray}
\label{tnn}
&&<\mbox{\boldmath$\varkappa$}^{*\prime}  \mu_1^\prime \mu_2^\prime |t_{c.m.}|
\mbox{\boldmath$\varkappa$}^* \mu_1\mu_2>
=<\mbox{\boldmath$\varkappa$}^{*\prime}  \mu_1^\prime \mu_2^\prime |
A+B(\mbox{\boldmath$\sigma_1$} \hat N^*)(\mbox{\boldmath$\sigma_2$} \hat N^*)+
\\
&&
C(\mbox{\boldmath$\sigma_1$} +\mbox{\boldmath$\sigma_2$} )\cdot \hat N^* +
D(\mbox{\boldmath$\sigma_1$} \hat q^*)(\mbox{\boldmath$\sigma_2$} \hat q^*) +
F(\mbox{\boldmath$\sigma_1$} \hat Q^*)(\mbox{\boldmath$\sigma_2$} \hat Q^*)
|\mbox{\boldmath$\varkappa$}^* \mu_1\mu_2>.
\nonumber
\end{eqnarray}
The orthonormal basis $\{\hat q^*,\hat Q^*,\hat N^*\}$ is a combination of the nucleon
relative momenta in the initial \mbox{\boldmath$\varkappa $}$^*$ and final 
\mbox{\boldmath$\varkappa$}$^{\prime *}$
states:
\begin{equation}
\hat q^*=\frac {\mbox{\boldmath $\varkappa$}^* -\mbox{\boldmath$\varkappa$}^{*\prime}}
{|\mbox{\boldmath$\varkappa$}^* -\mbox{\boldmath$\varkappa$}^{*\prime}|},~~
\hat Q^*=\frac {\mbox{\boldmath$\varkappa$}^* +\mbox{\boldmath$\varkappa$}^{*\prime} }
{|\mbox{\boldmath$\varkappa$}^* +\mbox{\boldmath$\varkappa$}^{*\prime}|},~~
\hat N^*=\frac {\mbox{\boldmath$\varkappa$}^*  \times 
\mbox{\boldmath$\varkappa$}^{*\prime} }{|\mbox{\boldmath$\varkappa$}^*
\times\mbox{\boldmath$\varkappa$}^{*\prime} |}.
\end{equation}
The amplitudes $A,B,C,D,F$ are the functions of the center-of-mass
energy and scattering angle. The radial parts of these amplitudes are taken 
as a sum of Yukawa terms. A new fit of the model parameters \cite{newlf} was done  
in accordance with the  phase-shift-analysis data SP07 \cite{said}.

Since the matrix elements are expressed
via the effective $NN$-interaction operators sandwiched
between the initial and final plane-wave states, this construction 
can be extended to the off-shell case allowing the initial and final
states to get the current values of  \mbox{\boldmath $\varkappa$} and
\mbox{\boldmath $\varkappa^\prime$}. Obviously, this extrapolation does 
not change the general spin structure.

The double scattering contribution (Fig.1c) is defined by a deuteron wave function and two
nucleon-nucleon $t$-matrices. Here, a nucleon from the initial deuteron is
scattered on the beam proton as it was in the single-scattering case. But then 
one of the scattered nucleons again interacts with another nucleon from the initial
deuteron. Thus, we have second NN vertex.  The final deuteron is formed by scattered
nucleons from the first and second NN-scatterings. 
 Also we have here three-nucleon propagator:  
\begin{eqnarray}
\label{ds}
&&{\cal J}_{DS}=\int d\vec{p_2}d\vec{p_3^\prime}
<-\vec P_d {\cal M}_d^\prime|\Omega_d^\dagger 
|-\vec P_d-\vec p_3^\prime~m_2^\prime,
\vec p_3^\prime~ m_3^\prime>
\\
&&<\vec p^\prime~ m^\prime,
 -\vec P_d-\vec p_3^\prime~m_2^\prime, \vec p_3^\prime~m_3^\prime|
\nonumber\\
&&\frac{ 
t_{3(NN)}^1(E^\prime) 
t_{2(NN)}^1(E)+
[t_{3(NN)}^1(E^\prime)+
t_{3(NN)}^0(E^\prime)]
[t_{2(NN)}^1(E)+
t_{2(NN)}^0(E)]/4}{E_d+E_p-E_1 -E_2-E_3^\prime +i\varepsilon}
\nonumber\\
&&|\vec p~m, \vec p_2~ m_2, \vec P_d -\vec p_2~m_3>
<\vec p_2~ m_2,~\vec P_d -\vec p_2~ m_3|\Omega_d|\vec P_d {\cal M}_d>.
\nonumber
\end{eqnarray}
The argument of the $NN$-matrix is defined
as the three-nucleon on-shell energy excluding the energy of the nucleon which does not
participate in the interaction:
\begin{eqnarray}
E=E_d+E_p-E_2,~~~~~~E^\prime=E_d+E_p-E^\prime _3.
\end{eqnarray}

The structure of the delta amplitude (Fig.1d) looks like the double-scattering one. But here we have
$NN\to \Delta N$ matrices instead the nucleon-nucleon matrices and $NN\Delta$-propagator instead three-nucleon one.
\begin{eqnarray}
\label{delta}
&&{\cal J}_{\Delta}=2\int d\vec{p_2}d\vec{p_3^\prime}dE_\Delta d\vec{p_\Delta}\delta(E_\Delta-\sqrt{\mu^2+\vec{p_\Delta^2}})
\delta(\vec p+\vec {P_d}-\vec{p_2}-\vec{p_3}-\vec{p_\Delta})_1<\frac{1}{2}\tau^\prime\frac{1}{2} m^\prime\vec{p^\prime}|
\nonumber\\
&&_{23}<00;\vec{-P_d}1{\cal M}_d^\prime |\Omega_d^\dagger[1-P_{12}]|
t_{3(N\Delta)}(E^\prime)\frac{1}{E-E_2-E_3^\prime-E_\Delta+i\Gamma(E_\Delta/2)}|\Psi_{\vec p_\Delta}(E_\Delta)>_1
\nonumber\\
&&
|\frac{1}{2}\tau_2\frac{1}{2}m_2\vec p_2;
\frac{1}{2}\tau_3\frac{1}{2}m_3\vec p_3>_{23}~_{23}<\frac{1}{2}\tau_2\frac{1}{2}m_2\vec p_2;
\frac{1}{2}\tau_3\frac{1}{2}m_3\vec p_3|
\\
&& _1<\Psi_{\vec p_\Delta}(E_\Delta)||t_{2(N\Delta)}(E)[1-P_{13}]\Omega_d|
\vec{P_d} 1{\cal M}_d; 00>_{23}|\frac{1}{2}\tau\frac{1}{2} m\vec{p}>_1
\nonumber
\end{eqnarray}

Here a full set of the particles quantum numbers was included into the amplitude definition.
Isospin and spin quantum numbers are marked by $\tau$ and $m$ or ${\cal M}_d$, respectively. The indexes
near the bracket correspond to the particle numbers. 

The distribution function of the delta energy
\begin{eqnarray}
|\Psi_{\vec p_\Delta}(E_\Delta)><\Psi_{\vec p_\Delta}(E_\Delta)|=\rho(E_\Delta)
\end{eqnarray}
is defined through the delta width $\Gamma(\mu)$:
\begin{eqnarray}
\rho(\mu)=\frac{1}{2\pi}\frac{\Gamma(\mu)}{(E_\Delta (\mu)-E_\Delta(m_\Delta))^2+\Gamma^2(\mu)/4},
\end{eqnarray}
where $\mu^2=E_\Delta^2-\vec p^2_\Delta$ is the squared four-momentum of the delta.
The delta width depends on the energy . Here we use  the standard parameterization of $\Gamma(\mu)$ taking into account
the $\Delta$ off-shell corrections \cite{jain},\cite{dmitr}:
\begin{eqnarray}
\Gamma(\mu)=\Gamma_0\frac{p^3(\mu^2,m_\pi^2)}{p^3(m^2_\Delta,m_\pi^2)}\cdot\frac{p^2(m^2_\Delta,m_\pi^2)+\gamma^2}{p^2(\mu^2,m_\pi^2)+\gamma^2}.
\end{eqnarray}

\vspace{0.5cm}\noindent
where $p(x^2,m_\pi^2)$ is the momentum in the $\pi N$ -center-of-mass:
\begin{eqnarray}
p(x^2,m_\pi^2)=\sqrt{(x^2+m_N^2-m_\pi^2)^2/4x^2-m_N^2}.
\end{eqnarray}
In our calculation, we use the following values of constants:
\begin{eqnarray} 
\Gamma_0=0.120 ~GeV,~~~~\gamma=0.200 ~GeV,~~~~m_\Delta=1.232 
\end{eqnarray}

In the  Born approximation, the $NN\to N\Delta$ $t$-matrix  can be replaced with the
corresponding potential:
\begin{eqnarray}
  <\vec p,\frac{1}{2}m,\frac{1}{2}\tau|t_{(N\Delta)}(E)|\Psi_{\vec p_\Delta}(E_\Delta)>\approx
 <\vec p,\frac{1}{2}m,\frac{1}{2}\tau|V_{(N\Delta)}(E)|\Psi_{\vec p_\Delta}(E_\Delta)>
\end{eqnarray}

The potential for the $NN\to N\Delta$ transition is based on  the $\pi-$ and $\rho-$ exchanges \cite{jain}:
\begin{eqnarray}
V_{\beta\alpha}^{(\pi)}&=&-\frac{f_\pi f_\pi^*}{m_\pi^2}F_\pi^2(t)\frac{q^2}{m_\pi^2-t}
(\vec\sigma\cdot\hat q)(\vec S\cdot \hat q)(\vec\tau\cdot\vec T)
\\
V_{\beta\alpha}^{(\rho)}&=&-\frac{f_\rho f_\rho^*}{m_\rho^2}F_\rho^2(t)\frac{q^2}{m_\rho^2-t}
\{(\vec\sigma\vec S)-(\vec\sigma\cdot\hat q)(\vec S\cdot \hat q)\}(\vec\tau\cdot\vec T)
\nonumber
\end{eqnarray}
Here, $t$ is the four transfer momentum and $\vec q$ is the corresponding three transfer momentum.
The  operators $\vec\sigma (\vec \tau)$  are $\frac{1}{2}$- spin (isospin) operators defined by Pauli
matrices while $\vec S (\vec T)$ operators correspond to $\frac{1}{2}\to \frac{3}{2}$ spin (isospin) transition.
$m_\pi$ and $m_\rho$ are  pion and $\rho$- meson masses. 
The coupling constant $f_\pi$ is related with the $NN\pi$ vertex and $f_\pi^*$ corresponds to the 
$N\Delta\pi$ one. It  concerns also $\rho-$ coupling constants.
\begin{eqnarray}
f_\pi&=&1.008~~~~~~f_\pi^*=2.156
\\
f_\rho&=&7.8~~~~~~f_\rho^*=1.85f_\rho
\nonumber
\end{eqnarray}
The hadronic form factor was chosen in a pole form as it was suggested in \cite{mach}:  
\begin{eqnarray}
F_x(t)=\left[(\Lambda_x^2-m_x^2)/(\Lambda_x^2-t)\right]^n
\end{eqnarray}
In our calculation, we use $\Lambda_\pi=0.8$ GeV, $\Lambda_\rho=1.8$ GeV. The exponent $n$ is equal to 1 for $\pi$-meson
and 2 for $\rho$-meson.

Since two nucleon states in the $NN\to N\Delta$ vertexes are antisymmetrized, two permutation operators appear in Eq.(\ref {delta}).
As consequence, the $\Delta$- amplitude contains four terms: one direct, two exchange, and one double-exchange ones.  
The permutation operator $P_{ij}$ involves the permutation of all quantum numbers. Here, it is permutation over momentum, spin, and
isospin indexes:   $P_{ij}=P_{ij}(p)P_{ij}(\sigma)P_{ij}(\tau)$.

%


\section{Polarization observables}
\label{sec:kin}
The $dp$ elastic scattering amplitude ${\cal J}_{dp\to dp}$ 
can be decomposed in spin-1 operators $S_i,~ Q_{ij}$ and spin-1/2 operators $\sigma_i$ acting in the deuteron and proton
spin space,respectively, as\cite{dp}: 
\begin{eqnarray}
\label{dp}
{\cal J}_{dp\to dp}&=&<m^\prime {\cal M}_d^\prime |f_1+f_2(\vec S\vec y) +
f_3 Q_{xx}+f_4 Q_{yy}+f_5 (\vec\sigma\vec x)(\vec S\vec x)+
\nonumber\\
&&
\hspace{1.7cm}
f_6(\vec\sigma\vec x) Q_{xy}+
f_7(\vec\sigma\vec y)+
f_8(\vec\sigma\vec y)(\vec S\vec y)+f_9 (\vec\sigma\vec y)Q_{xx}+
\\
&&\hspace{1.7cm}
f_{10}(\vec\sigma\vec y) Q_{yy}+
f_{11}(\vec\sigma\vec z)(\vec S\vec z)+
f_{12}(\vec\sigma\vec z)Q_{yz}|m {\cal M}_d>~~.
\nonumber
\end{eqnarray}
This decomposition is general and does not depend on the method by which the amplitude is calculated.
 Due to  the time-reversal and parity invariance, we have only 12 linearly 
 independent amplitudes $f_i$. 
The operators $\sigma_i$ are the Pauli
 matrices, while  $S_i$   and $Q_{ij}$ are the
 spin-1 vector and quadrupole operators, respectively:
\begin{eqnarray}
\label{Qij}
Q_{ij}=\frac{3}{2}(S_i S_j+S_j S_i)-2\delta_{ij} \hat I~,
~~~~~~~~
Q_{xx}+Q_{yy}+Q_{zz}=0~~.
\end{eqnarray} 
However, explicit expressions of the amplitudes can only be obtained  using some model.
In particular, the amplitudes $f_i$  can be calculated 
within the framework of the technique presented in the previous section. Note, if we know reaction amplitude
we can construct both an unpolarized differential cross section and
 any polarization observables.

The differential cross section of the  $dp$ elastic scattering is expressed in the center-of-mass via the squared reaction amplitude  by:
\begin{equation}
\frac {d\sigma}{d\Omega ^*}=(2\pi)^4\cdot\frac{1}{6}\cdot\frac{1}{s}Tr({\cal F F}^\dagger),
\end{equation}
where $s$ is the  Mandelstam invariant variable, $s=(P_d+p)^2$.
The invariant amplitude ${\cal F}$ is related to the amplitude in the Breit frame (or any other frame) as:
\begin{eqnarray}
{\cal F}=\sqrt{E_d^\prime E_p^\prime}~~{\cal J} \sqrt{E_d E_p}.
\nonumber
\end{eqnarray} 

The squared $dp$-amplitude summarized over all the spin projections can be presented
through amplitudes $f_i$ as 
\begin{eqnarray}
Tr({\cal J J}^\dagger)&=&6 (f_1^2+f_7^2)+4 (f_2^2+f_5^2+f_8^2+f_{11}^2)+
\frac{4}{3} (f_3^2+f_4^2+f_9^2+f_{10}^2)+
\\
&&
(f_6^2+f_{12}^2)-
\frac{4}{3} Re(f_3 f_4^*+f_9f_{10}^*)~~.
\nonumber
\end{eqnarray}

In this paper, we  consider only two polarization observables at the scattering angle in the center-of-mass
$\theta^*=180^\circ$.
The first of them is the tensor analyzing power $T_{20}$ and another is
 the polarization transfer from the initial deuteron to the final proton $\varkappa_0$. The observables are defined
through the reaction amplitude by relations \cite{ohlsen}: 

\begin{eqnarray}
\label{pol}
T_{20}=\frac{1}{\sqrt{2}}A_{zz}~,~~~~~~
A_{ij}=\frac{Tr({\cal J} Q_{ij} {\cal J}^\dagger)}{Tr({\cal J J}^\dagger)}~,
~~~~~~
\varkappa_0=\frac{3}{2}\frac{Tr({\cal J} S_y {\cal J}^\dagger\sigma_y)}{Tr({\cal J J}^\dagger)}
\end{eqnarray}
 
Due to Eq.(\ref{Qij}), $A_{zz}=-(A_{xx}+A_{yy})$. Moreover, $A_{xx}=A_{yy}$ at  $\theta^*=180^\circ$. In such a way,
it is enough to know the expression for $A_{yy}$ to find $T_{20}(180^\circ)$.
Using the standard technique and definitions (\ref{dp}), (\ref{pol}) one can calculate 
traces to define
 tensor analyzing powers
\begin{eqnarray}
Tr({\cal J} Q_{yy} {\cal J}^\dagger)&=&
4(f_2^2+f_8^2)+\frac{2}{3}(f_3^2+f_9^2)-
\frac{4}{3}(f_4^2+f_{10}^2)
-2(f_5^2+f_{11}^2)-
\nonumber\\
&&\frac{1}{2}(f_6^2+f_{12}^2)+
8 Re (f_1 f_4^*+f_7 f_{10}^*)-
4 Re (f_1 f_3^*+f_7 f_9^*)+
\\
&&\frac{4}{3} Re (f_3 f_4^*+ f_9 f_{10}^*)~~,
\nonumber
\end{eqnarray}
and polarization transfer from the deuteron to the proton
\begin{eqnarray}
Tr({\cal J} S_y  {\cal J}^\dagger\sigma_y)&=&
8 Re(f_1 f_8^*+f_2 f_7^*)-\frac{4}{3} Re(f_2 f_9^*+f_3 f_8^*)+
\frac{8}{3} Re(f_2 f_{10}^*+f_4 f_8^*)+
\nonumber\\
&&4 Re(f_5 f_{11}^*)+Re(f_6 f_{12}^*).
\end{eqnarray}
In general, we have to apply the spin transformation 
to relate the  observables in the center-of-mass and  Breit frames.
However, we have considered only those observables that are due to
 the polarization along the normal to the scattering plane.
In this case, the definitions of the polarization observables
are identical for both frames. This is not
the case for other observables such as $A_{xx}$, $A_{xz}$, etc.

\section{Results}
\label{sec:results}

We applied the method to describe angular dependence of the differential
cross sections at the backward scattering angles $\theta^*\ge 140^\circ$ at four deuteron energies
of 880, 1000, 1200, and 1300 MeV. All calculations were performed with CD Bonn deuteron wave function \cite{cd}.
 The  results are presented in Figs.2-5.  Four curves in the figures correspond
 to the calculations taking into account different reaction mechanisms.

It is well known that the data on the differential cross sections show
some enhancement at the backward angles what is well seen in the figures. However, the calculation results
obtained without the $\Delta$-isobar lie below the data. Moreover the difference
between the data and the results increases with the
 energy growing. When we include 
the $\Delta$-isobar into consideration we get rather good agreement between the data and
theory in the case when both $\pi$- and $\rho$-mesons are taken into account.
When we include only  $\pi$-meson in the  $\Delta$-isobar description we get
overestimated values for the differential cross sections. 
 Despite the fact that the inclusion
of the DS-term does not describe the rise, this term makes a significant contribution into the amplitude
and increases the value several times compared to the result obtained with  the ONE+SS term only.

It is interesting to look at the manifestation of the various mechanisms at the special
scattering angle of $180^\circ$. An energy dependence of the differential cross section is
presented in Fig.6. The data  demonstrate a shoulder at the energies between about 500 and 1400 MeV
which is not described by the calculations without $\Delta$-isobar. The curves obtained taking into account 
only ONE+SS and ONE+SS+DS rapidly descend
 and pass below the data. Inclusion of $\Delta$-isobar into consideration
allows us to describe the shoulder and significantly improve an agreement between the data and theoretical
predictions. As in the case of the angular distributions of the differential cross sections, we get overestimated
 values when we include only $\pi$- meson in the $\Delta$-isobar definition. 

Tensor analyzing power $T_{20}$ is presented in Fig.7 as a function of the deuteron energy. The result, which takes into account only the simplest
reaction mechanism -ONE, is close to the data obtained at low energies up to about 500 MeV. At higher energies, the data go up while
ONE curve drops to a minimum equal to $-\sqrt{2}$ at the deuteron energy of about 1 GeV. Adding the single- and 
double- scattering terms does not significantly improve the situation.  
The results obtained with $\Delta$-isobar are close to the data at the energy exceeding 1.3 GeV.
However,  in the energy range
between 0.7 and 1.3 GeV, the experimental data show a rise, while the curve rises slightly. Although, in this case,  we see a discrepancy
between the data and theory predictions, the behaviour of the $\Delta$-curve qualitatively corresponds  to the data.

The role of the $\Delta$-isobar is clearly manifested in the polarization transfer $\varkappa_0$, 
which is shown in Fig. 8 versus the deuteron energy. The data show  a sharp drop at the deuteron energy
between 400 and 600 MeV,  and then  reach a plateau. 
The results  obtained with the inclusion of  $\Delta$-isobar  reproduce
the shape of the data while the  results of the calculations performed without $\Delta$-isobar are far from the data.

\includegraphics[width=1.1\linewidth]{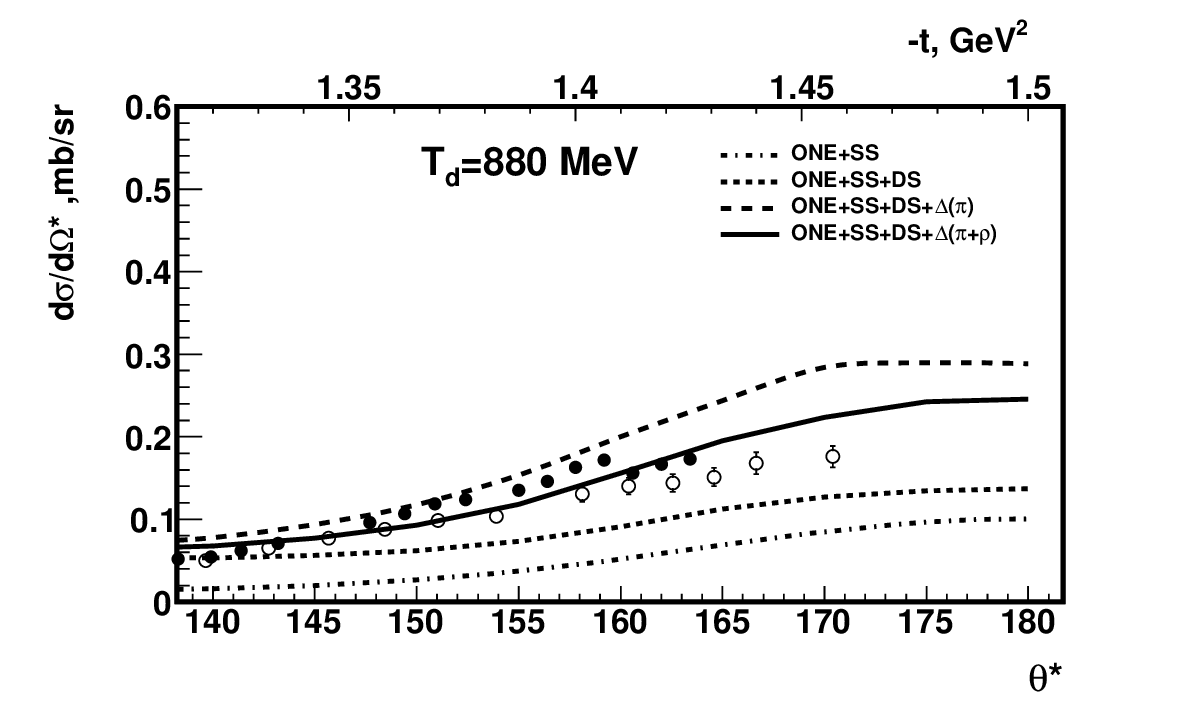}
{\footnotesize{Figure 2: The angular dependence of the differential cross section at
the deuteron energy $T_d=880$ MeV. The data are from  $\circ$ -\cite{cr880_a} at $T_d=850$ MeV
  $\bullet$-\cite{alder} at $T_d=940$ MeV}}

\includegraphics[width=1.1\linewidth]{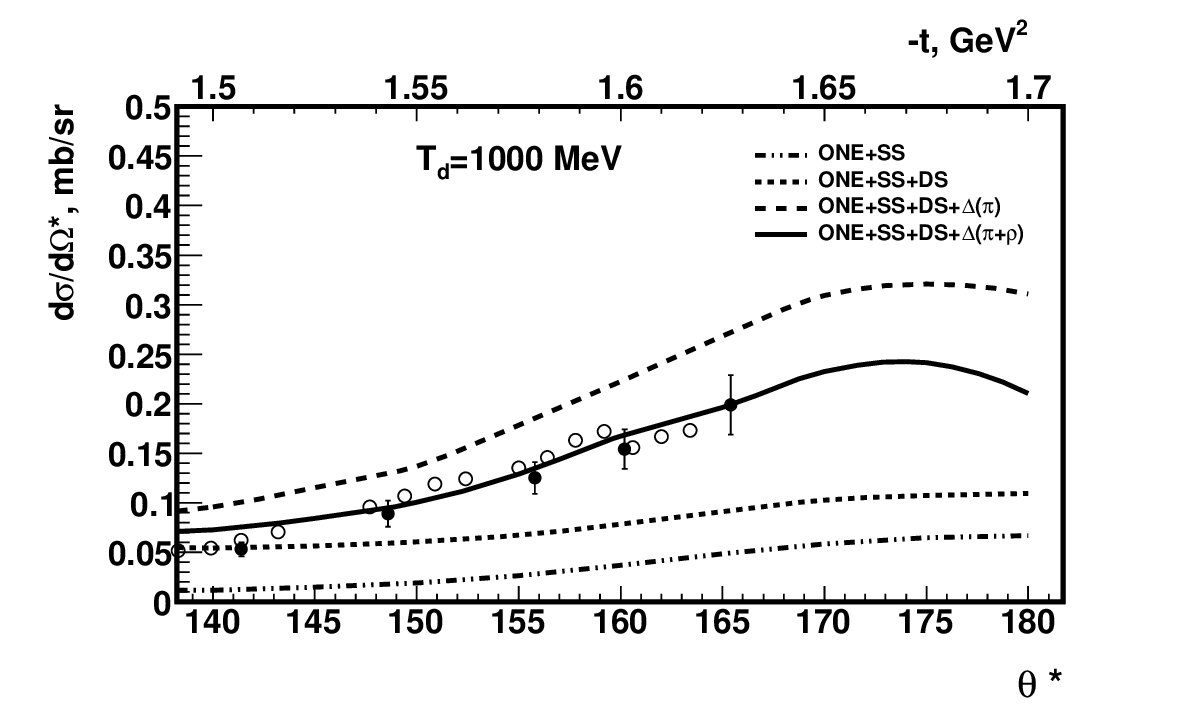}
{\footnotesize{Figure 3: The angular dependence of the differential cross section at
the deuteron energy at $T_d=1000$ MeV. The data are from   $\circ$-\cite{alder} at $T_d=940$ MeV, $\bullet$-\cite{vincent} at $T_d=1169$ MeV.
}}
\label{1000}       

\includegraphics[width=1.1\linewidth]{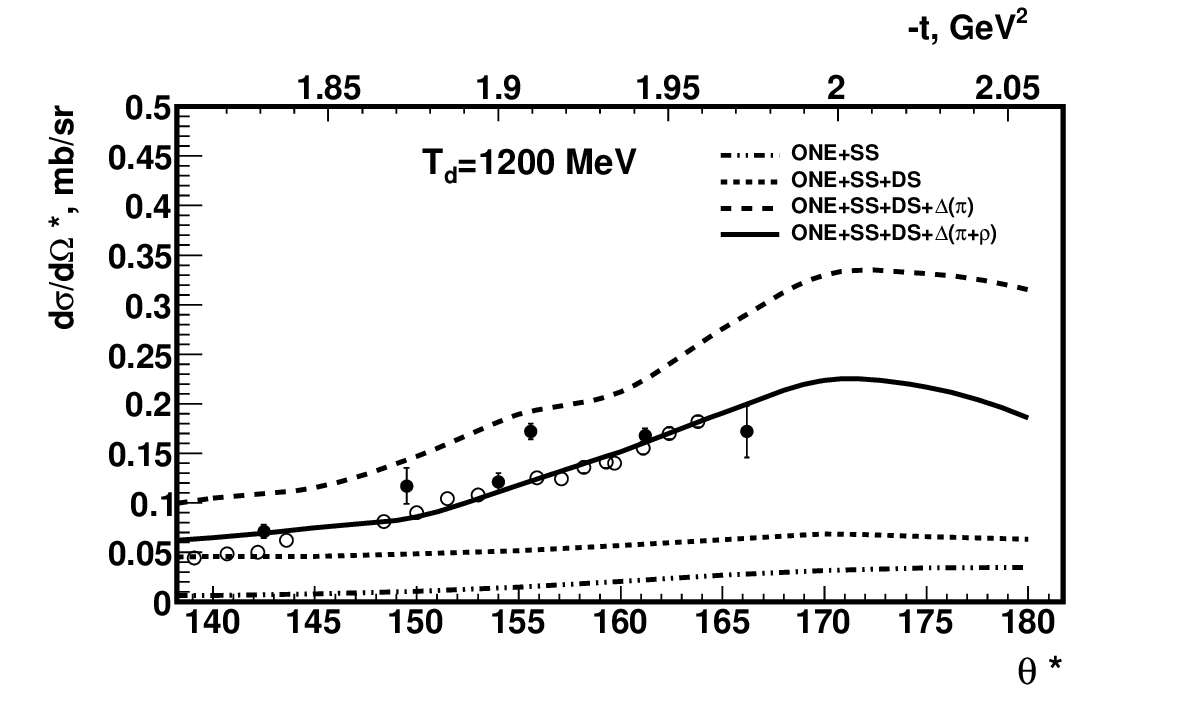}
{\footnotesize{Figure 4: The angular dependence of the differential cross section at
the deuteron energy at $T_d=1200$ MeV. The data are from $\circ$ -\cite{alder} at $T_d=1169$ MeV, $\bullet$ -\cite{cr1200} at $T_d=1200$ MeV. 
}}
\label{1200}       

\includegraphics[width=1.1\linewidth]{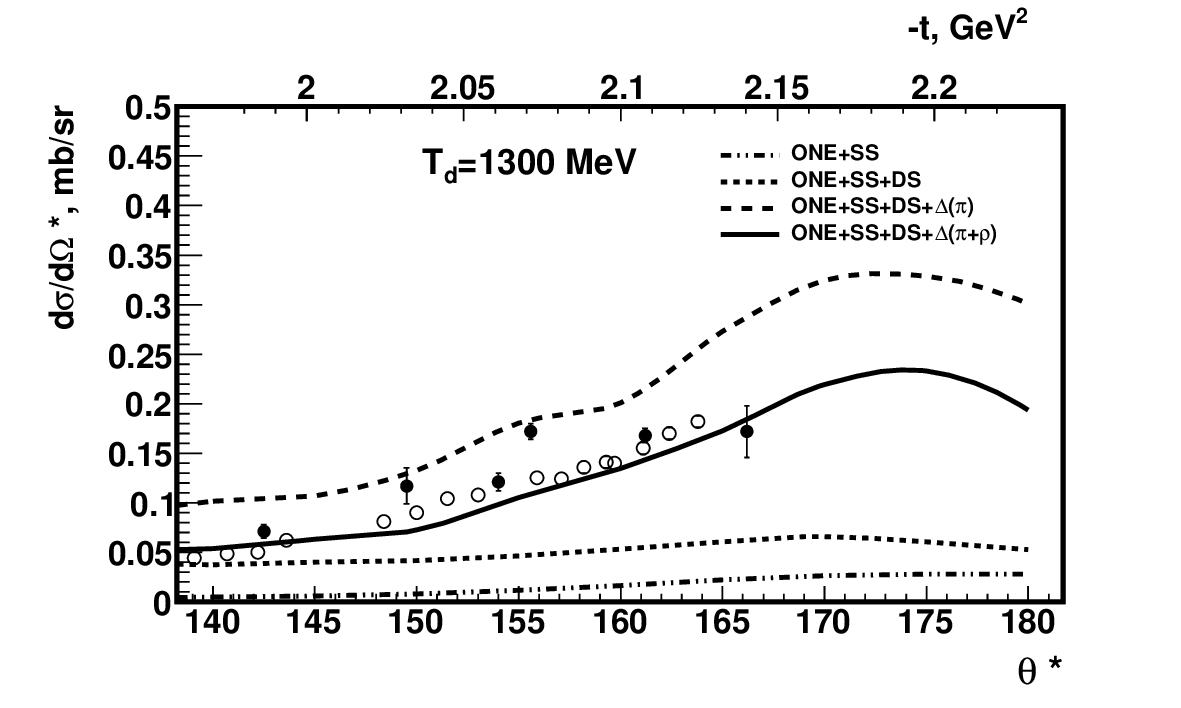}
{\footnotesize{Figure 5: The angular dependence of the differential cross section at
the deuteron energy $T_d=1300$ MeV. The data are from $\circ$ -\cite{alder} at $T_d=1169$ MeV, $\bullet$ -\cite{cr1200} at $T_d=1200$ MeV.
}}
\label{1300}       

\includegraphics[width=\linewidth]{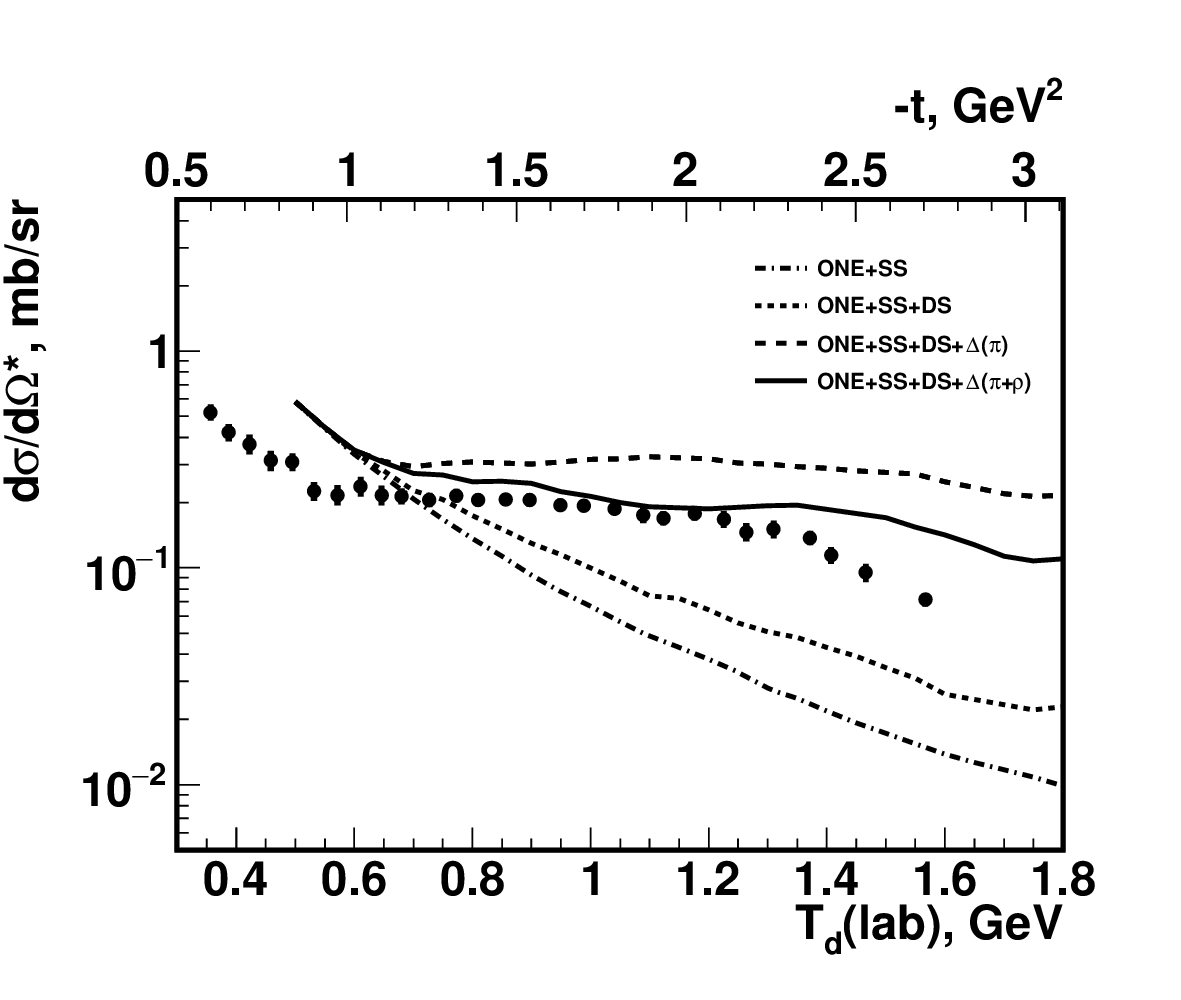}
{\footnotesize{Figure 6: The energy dependence of the differential cross section at
the scattering angle $\theta^*=180^\circ$. The data are from \cite{bonner}.
}}

\includegraphics[width=1.1\linewidth]{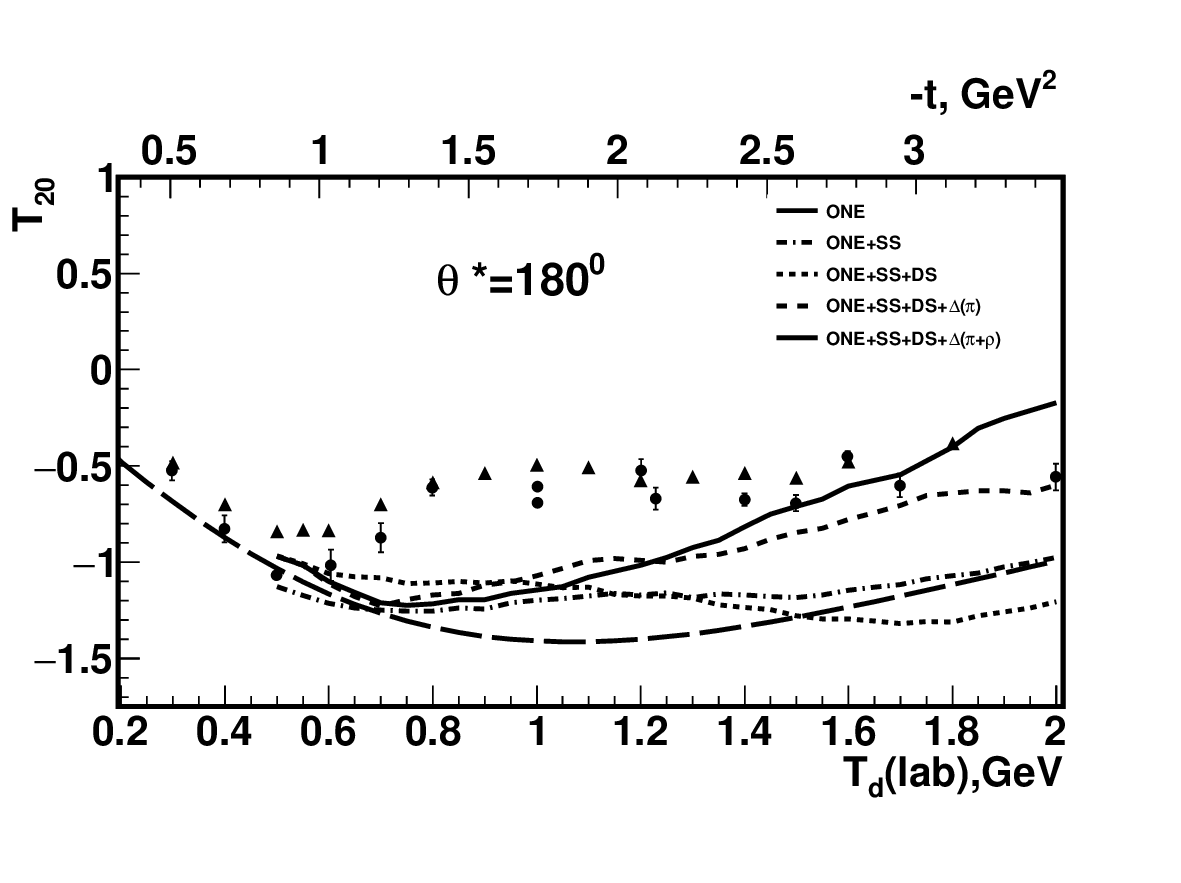}
{\footnotesize{Figure 7: The energy dependence of the differential cross section at
the scattering angle $\theta^*=180^\circ$. The data are taken from  $\bullet$ -\cite{saclay}, {$\blacktriangle$} -\cite{punjabi}. 
}}
\label{1200}       

\includegraphics[width=1.1\linewidth]{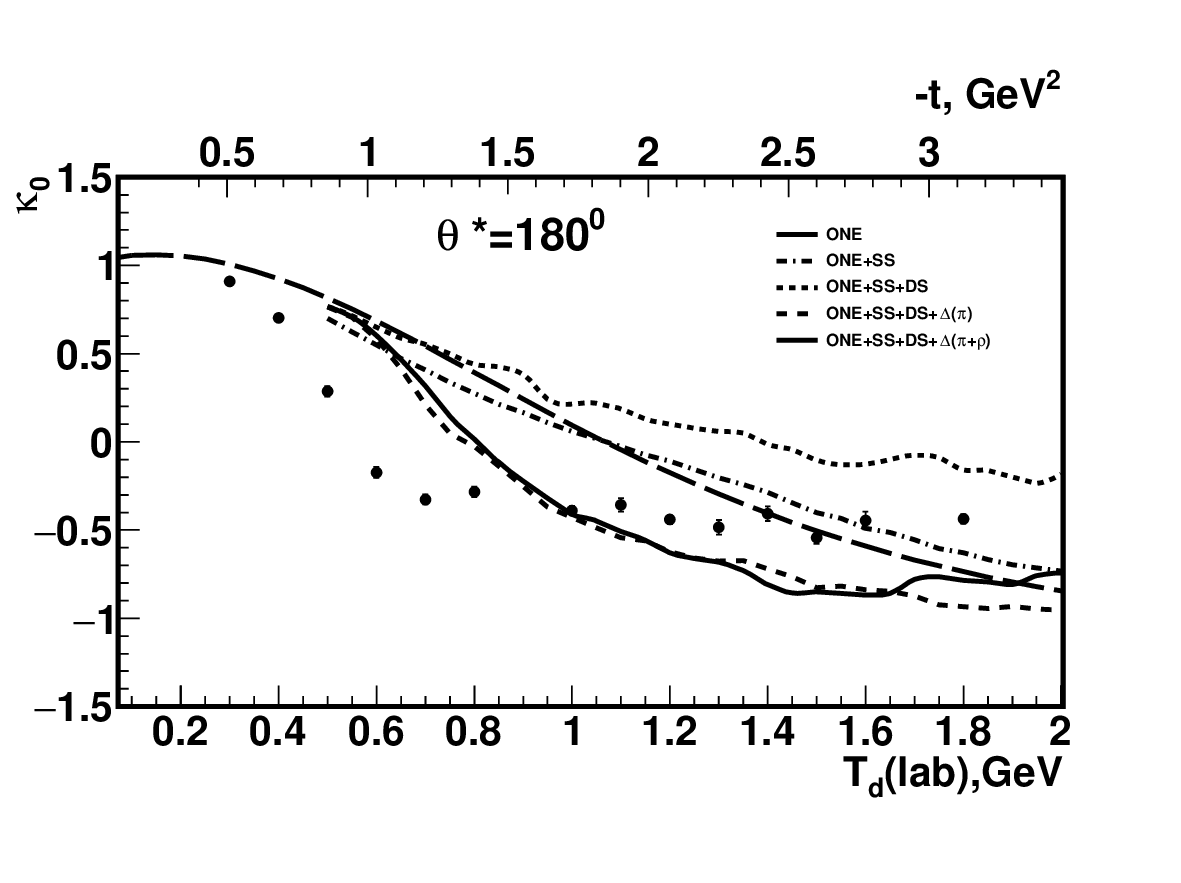}
{\footnotesize{Figure 8: The energy dependence of the polarization transfer at
the scattering angle $\theta^*=180^\circ$. The data are taken from \cite{punjabi}.
}}
\label{1300}       

\vspace{0.5cm}\noindent
\vspace{0.5cm}\noindent

\section{Conclusion}

We  considered four possible  mechanisms  of  the deuteron-proton backward
elastic scattering in GeV- energy range and 
 investigated the role of each of them. 
 The multiple-scattering   model was used  
  to calculate the reaction amplitude. The angular dependence of
the differential cross section has been obtained at $\theta^*\ge 140^\circ$ in the energy range between 880
and 1300 MeV. 
It has been shown that 
inclusion of the $\Delta$-isobar term into consideration allowed describing the  enhancement
of the differential cross section under these kinematical conditions.

We also studied the special case of the scattering angle $\theta^*= 180^\circ$.
A quite good agreement has been obtained  between the experimental data and the theoretical predictions
for the energy dependence of the differential cross section. We also considered  two polarization observables:
the tensor analyzing power $T_{20}$ 	and polarization transfer $\varkappa_0$.
As in the case of the differential cross section, the acceptable behaviour of the $T_{20}$ and $\varkappa_0$ curves was  achieved only after taking $\Delta$-isobar contribution into account.

In such a way, we can conclude that the mechanism when the $\Delta$-isobar  appears  in the intermediate state is decisive at the large angles. Nevertheless, the SS- and DS- mechanisms make a significant contribution into the reaction
amplitude, and they should not be neglected.

\section*{Acknowledgments}
The author is grateful to Dr. V.P. Ladygin for fruitful discussions and interest in this problem.

\end{document}